\documentclass[aps,prl,superscriptaddress,citeautoscript,twocolumn]{revtex4-2}

\usepackage{graphicx}
\usepackage{dcolumn}
\usepackage{amsmath,mathrsfs,amssymb,amsthm}
\usepackage{hhline}
\usepackage{wasysym,enumerate,color}
\usepackage{epstopdf}
\usepackage{verbatim}
\usepackage{hyperref}
\usepackage{color}
\usepackage{ulem} 

\newcommand{\bea}{\begin{eqnarray*}}
\newcommand{\eea}{\end{eqnarray*}}
\newcommand{\bne}{\begin{equation*}}
\newcommand{\ede}{\end{equation*}}

\newcommand{\bnen}{\begin{equation}}
\newcommand{\eden}{\end{equation}}
\newcommand{\bean}{\begin{eqnarray}}
\newcommand{\eean}{\end{eqnarray}}
\newcommand{\bnsn}{\begin{subequations}}
\newcommand{\edsn}{\end{subequations}}

\newcommand{\bna}{\begin{array}}
\newcommand{\eda}{\end{array}}
\newcommand{\bnm}{\begin{enumerate}}
\newcommand{\edm}{\end{enumerate}}

\definecolor{darkgreen}{rgb}{0,0.5,0}
\definecolor{purple}{rgb}{0.35,0,0.35}
\definecolor{orange}{rgb}{1,0.5,0}
\definecolor{darkred}{rgb}{.7,0,0}
\definecolor{darkblue}{rgb}{0,0,.3}
\definecolor{grey}{rgb}{.6,.6,.6}
\definecolor{dimgreen}{rgb}{0.2,0.6,0.1}
\hypersetup{colorlinks,linkcolor=blue,urlcolor=blue,citecolor=blue}

\begin{document}

\title{Multimode operation of a superconducting nanowire switch in the nanosecond regime}

\author{Zolt\'an~Scher\"ubl}
\affiliation{Department of Physics, Institute of Physics, Budapest University of Technology and Economics, M\H{u}egyetem rkp. 3., H-1111 Budapest, Hungary}
\affiliation{MTA-BME Superconducting Nanoelectronics Momentum Research Group, M\H{u}egyetem rkp. 3., H-1111 Budapest, Hungary}
\author{M\'aty\'as~Kocsis}
\affiliation{Department of Physics, Institute of Physics, Budapest University of Technology and Economics, M\H{u}egyetem rkp. 3., H-1111 Budapest, Hungary}
\affiliation{MTA-BME Superconducting Nanoelectronics Momentum Research Group, M\H{u}egyetem rkp. 3., H-1111 Budapest, Hungary}
\author{Tosson~Elalaily}
\affiliation{Department of Physics, Institute of Physics, Budapest University of Technology and Economics, M\H{u}egyetem rkp. 3., H-1111 Budapest, Hungary}
\affiliation{MTA-BME Superconducting Nanoelectronics Momentum Research Group, M\H{u}egyetem rkp. 3., H-1111 Budapest, Hungary}
\affiliation{Low-Temperature Laboratory, Department of Applied Physics, Aalto University School of Science, P.O. Box 15100, FI-00076, Aalto, Finland}
\author{L\H{o}rinc~Kup\'as}
\affiliation{Department of Physics, Institute of Physics, Budapest University of Technology and Economics, M\H{u}egyetem rkp. 3., H-1111 Budapest, Hungary}
\affiliation{MTA-BME Superconducting Nanoelectronics Momentum Research Group, M\H{u}egyetem rkp. 3., H-1111 Budapest, Hungary}
\author{Martin~Berke}
\affiliation{Department of Physics, Institute of Physics, Budapest University of Technology and Economics, M\H{u}egyetem rkp. 3., H-1111 Budapest, Hungary}
\affiliation{MTA-BME Superconducting Nanoelectronics Momentum Research Group, M\H{u}egyetem rkp. 3., H-1111 Budapest, Hungary}
\author{Gerg\H{o}~F\"ul\"op}
\affiliation{Department of Physics, Institute of Physics, Budapest University of Technology and Economics, M\H{u}egyetem rkp. 3., H-1111 Budapest, Hungary}
\affiliation{MTA-BME Superconducting Nanoelectronics Momentum Research Group, M\H{u}egyetem rkp. 3., H-1111 Budapest, Hungary}
\author{Thomas~Kanne}
\affiliation{Center for Quantum Devices and Nano-Science Center, Niels Bohr Institute, University of Copenhagen, Universitetsparken 5, DK-2100, Copenhagen, Denmark}
\author{Karl~Berggren}
\affiliation{Research Laboratory of Electronics, Massachusetts Institute of Technology, Cambridge, Massachusetts 02139, USA}
\author{Jesper~Nyg\r{a}rd}
\affiliation{Center for Quantum Devices and Nano-Science Center, Niels Bohr Institute, University of Copenhagen, Universitetsparken 5, DK-2100, Copenhagen, Denmark}
\author{Szabolcs~Csonka}
\email{csonka.szabolcs@ttk.bme.hu}
\affiliation{Department of Physics, Institute of Physics, Budapest University of Technology and Economics, M\H{u}egyetem rkp. 3., H-1111 Budapest, Hungary}
\affiliation{MTA-BME Superconducting Nanoelectronics Momentum Research Group, M\H{u}egyetem rkp. 3., H-1111 Budapest, Hungary}
\affiliation{HUN-REN Centre for Energy Research, H-1121 Budapest, Konkoly Thege Mikl\'os \'ut 29-33.}
\author{P\'eter~Makk}
\email{makk.peter@ttk.bme.hu}
\affiliation{Department of Physics, Institute of Physics, Budapest University of Technology and Economics, M\H{u}egyetem rkp. 3., H-1111 Budapest, Hungary}
\affiliation{MTA-BME Correlated van der Waals Structures Momentum Research Group, M\H{u}egyetem rkp. 3., H-1111 Budapest, Hungary}

\begin{abstract}
Superconducting circuits are promising candidates for future computational architectures, however, practical applications require fast operation.
Here, we demonstrate fast, gate-based switching of an Al nanowire-based superconducting switch in time-domain experiments. We apply voltage pulses on the gate while monitoring the microwave transmission of the device. Utilizing the usual leakage-based operation these measurements yield a fast, 1--2~ns switching time to the normal state, possibly limited by the bandwidth of our setup, and a 10--20~ns delay in the normal to superconducting transition. However, having a significant capacitance between the gate and the device allows for a novel operation, where the displacement current, induced by the fast gate pulses, drives the transition. The switching from superconducting to the normal state yields a similar fast timescale, while in the opposite direction the switching is significantly faster (4--6~ns) than the leakage based operation, which may be further improved by better thermal design. The measured short timescales and novel switching operation open the way for future fast and low-power-consumption applications. 
\end{abstract}

\date{\today}
\maketitle

\section{Introduction}
In 2018 De Simoni et al.\ demonstrated a novel operation of a superconducting circuit, where the switching current of a narrow, fully metallic wire was suppressed and eventually quenched by applying a DC voltage on a nearby gate electrode \cite{DeSimoniNatNano2018}. This effect was termed gate-controlled supercurrent (GCS), and since then it was investigated intensively \cite{PaolucciNanoLett2018,DeSimoniACSNano2019,PaolucciAVSQS2019,PaolucciPRAppl2019,PaolucciNanoLett2019,DeSimoniAPL2020,RocciACSNano2020,PugliaAPL2020,PugliaPRAppl2020,BoursPRResearch2020,OrusSciRep2021,AlegriaNatNano2021,SolinasPRL2021,RitterNatComm2021,GolokolenovNatComm2021,ElalailyNanoLett2021,BassetPRResearch2021,DeSimoniACSAEM2021,PaolucciNanoLett2021,MercaldoPRResearch2021,ChirolliPRResearch2021,CattoSciRep2022,RitterNatEl2022,AmorettiPRResearch2022,ElalailyACSNano2023,ChakrabortyPRB2023,DuSST2023,YuNanoLett2023,ZhangUnPub2023,KochNanoResearch2024,RyuNanoLett2024,ElalailyArXiv2024,JointArxiv2024,RufArXiv2024}.

Such an all electrically driven, fully metallic, superconducting switch could be an important building block of superconducting computing structures \cite{HolmesIEEETASC2013,SolovievBJNano2017,RufApplPhysRev2024}, since the superconducting state promises dissipationless operation, whereas the electrical control suggests a scalable architecture. However, for practical applications fast switching is required. Up to now two studies have investigated the switching speed and found $\sim 90$~ns \cite{RitterNatComm2021} and $\sim 25$~ns \cite{JointArxiv2024} timescales, which were limited by their setup in both cases.

In earlier works the GCS was explained by a purely electric-field-based effect \cite{DeSimoniNatNano2018,PaolucciPRAppl2019,PaolucciNanoLett2019,DeSimoniACSNano2019,BoursPRResearch2020,PugliaPRAppl2020,PugliaPRAppl2020,RocciACSNano2020,DeSimoniAPL2020,PaolucciNanoLett2021,OrusSciRep2021,DeSimoniACSAEM2021,YuNanoLett2023,KochNanoResearch2024}, but later works showed that the underlying mechanism in most cases is related to the leakage current in the substrate at high gate voltages \cite{AlegriaNatNano2021,RitterNatComm2021,GolokolenovNatComm2021,ElalailyNanoLett2021,BassetPRResearch2021,RitterNatEl2022,ElalailyACSNano2023,ElalailyArXiv2024} (see a detailed review of the field in Ref.~\cite{RufApplPhysRev2024}). Although the speed of a pure electric field based switching is expected to be only limited by the superconducting gap, yielding up to close to THz operation speed \cite{RufApplPhysRev2024}, but a leakage-based switching can be strongly limited by the large resistance of the leakage channel, stray capacitances and internal thermal timescales.

Aside from the leakage-current-based mechanisms, having a non-negligible capacitance, $C$ between the gate electrode and one of the contacts (see Fig.~\ref{fig:fig1}a) allows for a change of the gate voltage to induce a current in the lead, called the displacement current, $I_{\text{disp}}=C \cdot dV_{\text{g}}/dt$, which can switch the device to the normal state if it exceeds the switching current. This displacement-current-based architecture could be used for low-dissipation classical computing, pulse generation in neuromorphic applications \cite{DuNatComm2017,BerggrenNanotech2020,ChenNatComm2023,NishiokaCommEng2024} or combined with the leakage current could be operated as a switch in quantum electronic devices \cite{HolmesIEEETASC2013,SolovievBJNano2017,RufApplPhysRev2024}.

In this paper we determine the switching times corresponding to both the leakage- and the displacement-current-driven switching mechanisms by measuring time-resolved RF transmission of a superconducting switch while different pulses were applied on the gate electrode. We show that in both cases the operation is limited by the normal to supra transition, which is 10--20~ns for the leakage-based and 4--6~ns with the displacement-current-based operation. The switching to normal state is 1--2~ns for both mechanisms, presumably limited by the bandwidth of our setup. The former switching is probably limited by thermal timescales in the system which could be optimized by proper system design.

\section{Results and Discussion}

\begin{figure}[tb]
	\begin{center}
	\includegraphics[width=\columnwidth]{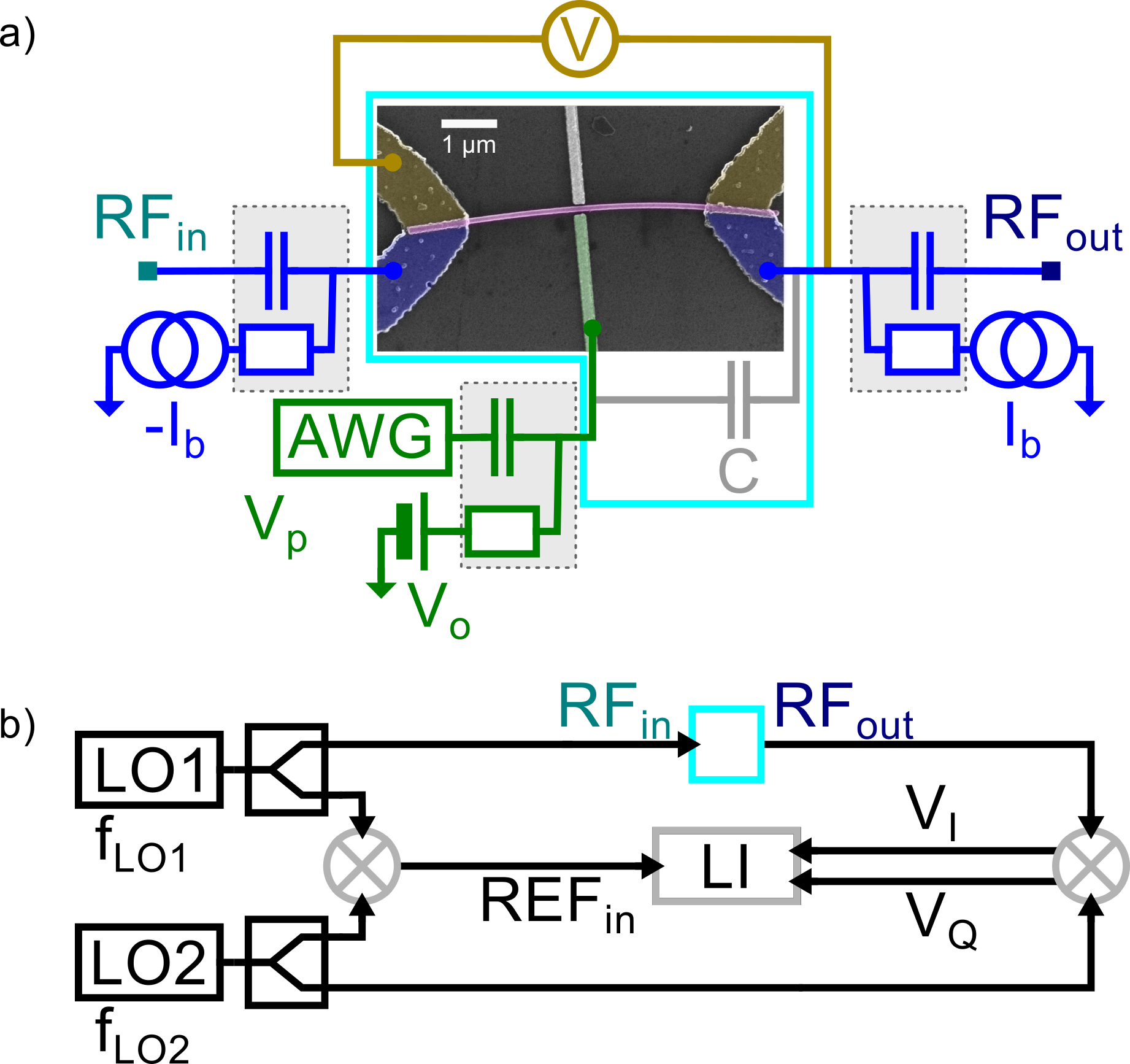}
	\caption{\textbf{The device and the measurement setup} a) the false-colored SEM image of the measured device along with the circuit diagram. The epitaxial Al shell of the InAs nanowire (purple) serves as the superconducting wire, contacted in a quasi-four-point geometry. Bias tees (gray rectangles) allow to combine DC and RF signals applied on the wire. The voltage drop across the wire is measured (yellow) to perform DC characterization. The side gate (green) is also connected to DC and RF sources to allow for fast pulsing. Due to the design (see Supp.\ Mat.\ I.), a substantial capacitance, $C$ (in gray) is formed between gate electrode and ohmic contacts. b) RF measurement setup. The cyan box corresponds to the same on panel a). The effect of the gate pulses was detected as the modulation of the transmission of a 3.95~GHz signal through the device, which was measured by a lock-in amplifier (LI) after downconversion using IQ mixers, hence both quadratures ($V_{\text{I}}$ and $V_{\text{Q}}$) of the signal were measured.}
	\label{fig:fig1}
	\end{center}
	\end{figure}

Our GCS device is based (see Fig.~\ref{fig:fig1}a) on an InAs nanowire with an epitaxial Al shell (purple) \cite{ElalailyNanoLett2021}, where the 20-nm-thin Al shell serves as the superconducting wire. A gate electrode (green) is fabricated 40~nm from the wire by electron beam lithography. The wire is contacted in quasi-four-point geometry by Ti/Al leads (yellow and blue). The GCS effect in this system was previously demonstrated in Refs.~\onlinecite{ElalailyNanoLett2021,ElalailyArXiv2024}. The current leads (blue) are coupled to both RF and DC lines through on-board bias tees (gray dashed rectangles). The device was symmetrically current-biased to keep its potential fixed during biasing. The voltage probe leads (yellow) are connected only to DC lines. The gate electrode (green) is also connected to a bias tee to combine the DC gate voltage with fast pulses or continuous wave excitations.

To investigate the device response to fast gate operations we have studied the time dependence of a high-frequency signal transmitted through the device, which was measured the following way (the RF measurement setup is illustrated in Fig.~\ref{fig:fig1}b). A fixed $f_{\text{LO1}}=3.95$~GHz signal was applied on one of the current leads of the device and the output signal was measured after downconversion (using IQ mixers) with a second local oscillator (LO2) with frequency $f_{\text{LO2}}$. Two different measurement schemes were used, \textit{technique A} (heterodyne): the transmitted signal was downconverted to a few hundred MHz (within the 600~MHz bandwidth of the Zurich Instruments (ZI) UHFLI, denoted by LI on Fig.~\ref{fig:fig1}b) and measured by lock-in technique and \textit{technique B} (homodyne): the measured signal was downconverted to DC with $f_{\text{LO2}}=f_{\text{LO1}}$ and the signal was directly digitized by the UHFLI. With technique A the reference signal was also generated by downconversion, while in technique B it was neglected. Further details of the device and the setup are given in the Methods and Supp.\ Mat.\ I. 

\begin{figure}[tb]
	\begin{center}
	\includegraphics[width=\columnwidth]{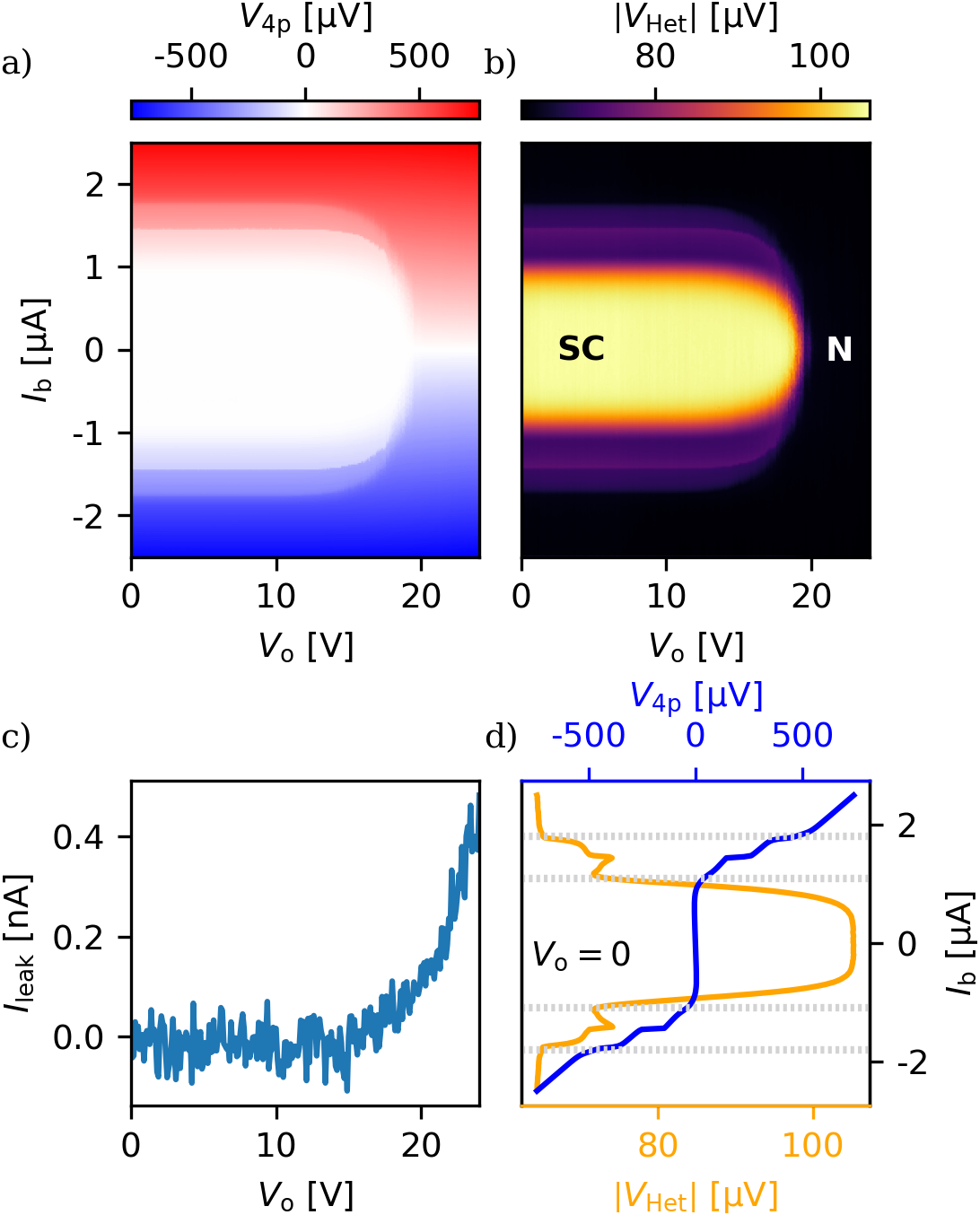}
	\caption{\textbf{The GCS effect and its RF measurement} a) The voltage drop, $V_{\text{4p}}$ on the nanowire as the function of the DC gate voltage and the DC bias current. The switching current starts to decrease at 15~V and vanishes at 19~V. b) The simultaneously measured transmitted 3.95~GHz RF signal showing a good correlation with the DC measurement. We find high transmission in the superconducting state and low in the normal one. c) The leakage current in the gate line, also measured simultaneously, increases along with the decreasing switching current. d) Zero-gate vertical line cuts of panels a) and b), an I-V curve and the corresponding RF transmission.} 
	\label{fig:fig2}
	\end{center}
	\end{figure}

Fig.~\ref{fig:fig2}a shows the typical GCS behavior, the reduction of the wire's switching current with increasing gate voltage, through the four-point voltage drop on the wire, $V_{\text{4p}}$ as a function of the DC gate voltage, $V_{\text{o}}$ and the DC bias current, $I_{\text{b}}$. Up to roughly 14~V the switching current is not affected by the gate voltage, above which the switching current starts to decrease and eventually vanishes at the threshold voltage, $V_{\text{th}}\approx 19$~V. At the same time the leakage current, measured on the gate line, is increasing exponentially (see panel c) \cite{ElalailyNanoLett2021,ElalailyACSNano2023,ElalailyArXiv2024}. Fig.~\ref{fig:fig2}b shows the simultaneously measured RF transmission, as the magnitude of the measured heterodyne voltage, $\left| V_{\text{Het}} \right|$, using technique A at $f_{\text{LO1}}=3.95$~GHz and $f_{\text{LO2}}=3.55$~GHz. The signal is the highest at low bias and low gate voltage, in the superconducting state (SC). Towards higher bias currents and gate voltages the transmission drops, following the increase of the DC resistance plotted on panel a). Further details are given in Supp.\ Mat.\ III. Fig.~\ref{fig:fig2}d shows the zero-gate vertical line cuts of panels a and b, further highlighting the good correspondence of the DC voltage curve and the RF transmission. The step-like voltage jumps, corresponding to the switching currents of different sections of the device \cite{ElalailyNanoLett2021}, appear as kinks in the mostly monotonically decreasing transmission. Altogether, our measurements demonstrate that the RF transmission is also a reliable tool to determine the state of the wire, with a much higher measurement speed compared to DC measurements.
	
First, we demonstrate the importance of the displacement current in our superconducting switch by investigating the device response to fast gate pulses using the following protocol. A DC offset gate voltage ($V_{\text{o}}$) was fixed, then a symmetric waveform with zero average, synthesized by an arbitrary waveform generator (ZI HDAWG, denoted by AWG on Fig.~\ref{fig:fig1}a) was applied on the RF input of the gate and at the same time the time-resolved RF transmission of the wire was recorded using technique A. To simulatenously have a good enough time resolution and a good signal-to-noise ratio, the outlined sequence is repeated and averaged 500-2000 times, where special care was taken to synchronize the pulses and the data acquisition. These measurements were carried out at zero DC current bias.

	\begin{figure}[tb]
	\begin{center}
	\includegraphics[width=\columnwidth]{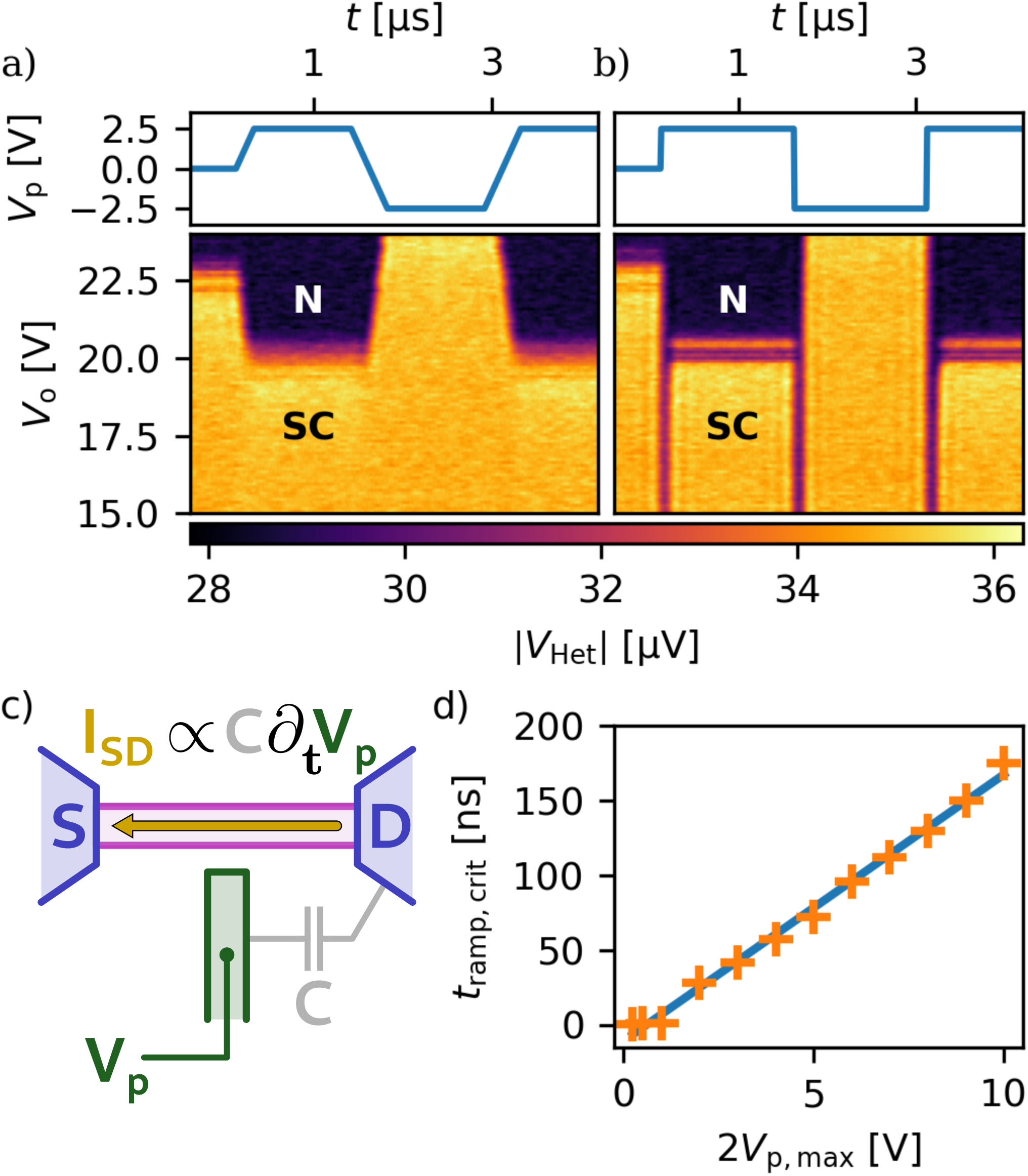}
	\caption{\textbf{Pulsed measurements} a) and b) Top panel: the applied pulse sequence, $V_{\text{p}}$, a) ramped, trapezoid-like pulse, b) square pulse. Bottom panels: the measured transmission (using technique A) as a function of time and the DC offset gate voltage. a) In case of a slowly ramped pulse the transmission follows well the total gate voltage, $V_{\text{o}}+V_{\text{p}}$, b) with the square pulse additional normal regions appear (vertical purple lines) each time the gate voltage suddenly changes, due to the generated displacement current and the resulting switching. c) Origin of the displacement current, being proportional to the capacitance between the gate and the device and gate voltage changing rate, d) The critical ramp times (see text for definition), which is needed for the switching, show a clear linear trend with the pulse amplitude, in agreement with the displacement current picture.}
	\label{fig:fig3}
	\end{center}
	\end{figure}

Two of such measurements are presented in Fig.~\ref{fig:fig3}a and b, where the top row shows the applied waveform, $V_{\text{p}}$, the bottom row shows the measured transmission, the magnitude of the heterodyne voltage, $\left| V_{\text{Het}} \right|$ as a function of time and the DC offset gate voltage, $V_{\text{o}}$. On panel a) a measurement is shown where a $\pm2.5$~V trapezoid-like wave was applied on the gate, with a slow, 200~ns ramp time (i.e. 25~V/\textmu s ramp rate). Taking a vertical cut of the measurements close to $t=0$ one obtains the suppression of superconductivity with DC gate voltage, already presented in Fig.~\ref{fig:fig2}a and b: with increasing gate voltage the wire switches from SC state (yellow) to normal (N) (purple - black). The threshold voltage differs from the one on Fig.~\ref{fig:fig1} due to the so-called training effect (see the Methods) \cite{ElalailyNanoLett2021}. At lower gate voltages, below 20~V, the wire remains in the SC state throughout the pulse sequence, since the total gate voltage ($V_{\text{g}}=V_{\text{o}}+V_{\text{p}}$) never exceeds the threshold voltage. However, for larger DC gate voltages, e.g. 21~V, a change in transmission is visible for the positive half of the pulse, where the total gate voltage (23.5~V) exceeds $V_{\text{th}}$ and hence the superconductivity is quenched (black region). On the negative side of the pulse the device switches back to the SC state (yellow). These results demonstrate that the state of the wire follows the applied pulse sequence and it is only determined by the total gate voltage. This suggests that wire state changes much faster than the ramp time of the pulse sequence and the device behaves in a quasi-stationary way.

Fig.~\ref{fig:fig3}b shows a similar measurement, however, using a square pulse (with ramp times limited by our instruments, $\sim 1$~ns). Compared to panel a) the boundaries between the SC and N regions become vertical, but the apparent threshold voltage again shifts with $-V_{\text{p}}$ as in the previous case. In addition, purple vertical lines appear both at the \textit{rising and falling edge} of the pulse. These features extend down to zero DC gate voltage (not shown here), but they do not extend into the normal region. Therefore, this effect has to be related to the state of the wire. This surprising feature means that independently from the DC gate voltage, the wire switches to the N state each time the gate voltage is \textit{rapidly} changed. After the rapid gate ramp the wire does not remain in the N state, but it switches back to the SC state if the total voltage does not exceed the threshold voltage. We attribute this switching effect to the displacement current in the wire and the leads, induced by the change of the gate voltage. Though the capacitance between the gate and the wire is rather small (30~aF), a substantial capacitance of 10--100~fF forms between the leads and the gate line on the millimeter-scale wiring segments from the device to the bonding pads (see the schematic in Fig.~\ref{fig:fig3}c and Supp.\ Mat.\ I.\ for the detailed discussion). Due to the asymmetric design of the device the gate-lead capacitance differs on the two sides of the nanowire, hence the displacement current results in a net current across the wire. If the induced displacement current is larger than the switching current, the wire indeed switches to N state in response to the fast change of the gate voltage. Since the displacement current is generated by the changing gate voltage, after the ramping period the displacement current decays and the wire switches back to SC state. Therefore, we call this effect displacement-current-induced switching (DCIS). 

Next, we investigated the effect of the ramping speed of the pulse on the DCIS. We repeated the measurement sequence with different pulse amplitudes and ramp times at zero DC gate voltage, but with improved time resolution, using technique B. We have already shown that if the ramp is slow enough, the DCIS is absent, so for a given pulse amplitude one can find a ramp time, below which the displacement current is large enough to switch the wire, but not above. The obtained \textit{critical ramp times}, $t_{\text{ramp,crit}}$ are shown in Fig.~\ref{fig:fig3}d as the function the gate pulse amplitude. Details of the evaluation are given in the Supp.\ Mat.\ V. The points indicate a clear linear trend, consistent with the displacement current picture, i.e., $I_{\text{disp}} = C_{\text{eff}} \cdot dV_{\text{g}}/dt$, where $C_{\text{eff}}$ is the effective capacitance between the gate and the device. From a linear fit, in blue, assuming that the displacement current is equal to the lowest switching current, $I_{\text{disp}} = I_{\text{c1}}=1.1~\mu$A, we found $C_{\text{eff}} = 20$~fF which is comparable with our finite element simulation (see Supp.\ Mat.\ I.\ for the details with a short discussion on the possible errors as well). 

\begin{figure}[tb]
	\begin{center}
	\includegraphics[width=\columnwidth]{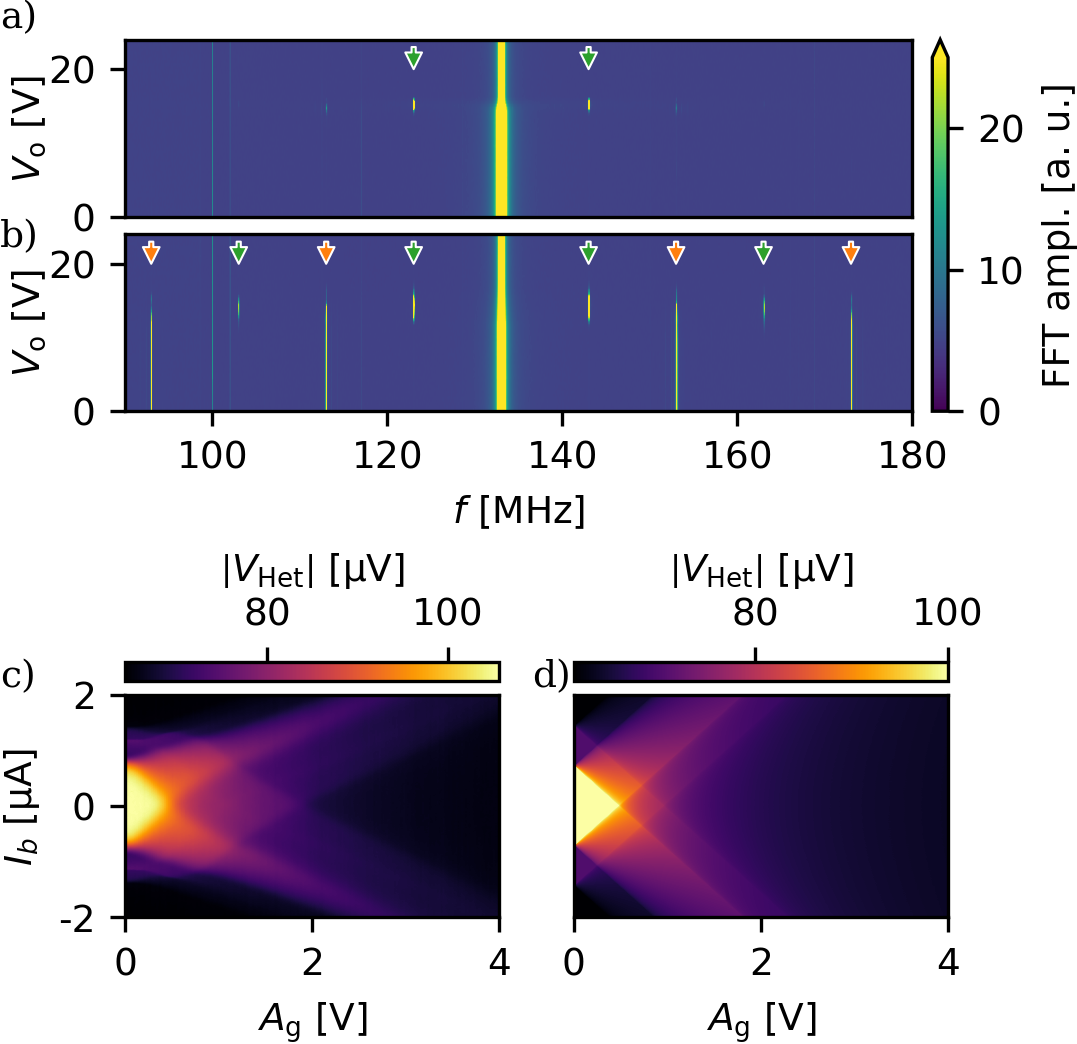}
	\caption{\textbf{Measurements with continuous harmonic drive.} a) and b) FFT spectra of the measured RF signal (after downconversion), while continuously driving the gate at $f_{\text{g}}=10$~MHz, with a) $A_{\text{g}}=0.5$~V, b) $A_{\text{g}}=2$~V amplitude as the function of the DC offset gate voltage, measured at zero DC bias current. The total gate voltage is $V_{\text{g}}=V_{\text{o}} + A_{\text{g}} \text{sin} \left( 2 \pi f_{\text{g}} t \right)$. The central peak at 133~MHz corresponds to the direct transmission of the readout signal. The sideband peaks are the result of mixing $f_{\text{g}}$ with the readout signal. The odd sideband peaks at 123 and 143~MHz around $V_{\text{o}}=15$~V, the threshold voltage, are due to the gating effect on panel a), while the even sideband peaks at 113 and 153~MHz, extending down to zero gate voltage on panel b) is the result of DCIS. c) Time-averaged transmission of the device with $f_{\text{g}}=15$~MHz harmonic drive as the function of the drive amplitude, $A_{\text{g}}$ and the DC bias current, $I_{\text{b}}$ at zero DC gate voltage. The superconducting region (yellow) shrinks with increasing drive amplitude. At large amplitudes better transmitting branches appear at high bias, where the displacement current is partially compensated by the DC bias. d) Simulation for panel c), where the instantaneous transmission is determined by the total current flowing through the device (the sum of the DC bias and the harmonic displacement current) and time-averaged for one period of the drive, using $I_{\text{c1}}=0.9$~\textmu A, $I_{\text{c2}}=1.8$~\textmu A, $C=20$~fF.}
	\label{fig:fig4}
	\end{center}
	\end{figure}

After these time-domain measurements we have turned to frequency-domain studies, where the response to a harmonic excitation can be addressed. Applying a continuous sine wave on the gate opens up the possibility to study mixing effects. The periodic modulation of the gate voltage also modulates the high frequency transmission (by changing the state of the wire) and mixes the signal on the gate with the readout tone. To investigate this effect the output signal was mixed down to 133~MHz (technique A with $f_{\text{LO1}}=3.95$~GHz and $f_{\text{LO2}}=3.817$~GHz) and an FFT was taken (using the scope module of the UHFLI). Two of such examples are shown in Fig.~\ref{fig:fig4}a and b, where the FFT amplitude is shown as the function of the offset gate voltage and the frequency. During these measurements a combined DC plus AC voltage was applied on the gate: $V_{\text{o}}+A_{\text{g}} \text{sin} \left(2\pi f_{\text{g}} t\right)$, with $A_{\text{g}}=0.5$~V for panel a) and 2~V for panel b) with $f_{\text{g}}=10$~MHz. For both measurements a large signal is visible at $f_{\text{LO1}}-f_{\text{LO2}}=133$~MHz in the full offset gate voltage range, coming from the measurement tone. For the smaller amplitude on panel a) other features show up in a narrow offset gate range from 14~V to 16~V at higher and lower frequencies than the measurement tone, marked by the green arrows. These side peaks originate from the mixing of the harmonic gate drive with the transmitted signal. The two dominant sideband peaks that are spaced $\pm f_{\text{g}}$ from the main peak at 123~MHZ and 143~MHz are indicated by the arrows. We interpret the presence of these peaks as the leakage-driven switching. When the DC offset plus the drive is larger than the threshold, the wire is switched to N state, but when the total gate voltage decreases below the threshold, the wire switches back to the SC state. Hence, the switching occurs once every period and the transmission is modulated at $f_{\text{g}}$. 

In sharp contrast, for $A_{\text{g}}=2.5$~V, four pairs of sideband peaks appear around the main peak (see panel b). The first and third sideband peaks (indicated by green arrows) only appear around the threshold gate voltage, while the second and fourth (orange arrows) extend down to zero DC gate voltage. The latter ones appear due to the displacement current, since for both increasing and decreasing gate voltage the displacement current exceeds the switching current. Therefore, the wire switches to the N state \textit{twice} in one period of the gate drive, resulting in a frequency doubling. These measurements provide a clear distinction of the leakage-driven and displacement-current-based switching mechanisms. As a control experiment, at large offset gate voltage, in the N state no sideband could be observed. We also measured the mixing effect at several drive frequencies and amplitudes. Using a simple model we could well reproduce the general tendencies as we demonstrate in Supp.\ Mat.\ VII-VIII.

The DCIS picture can be further validated by applying a DC bias current that adds to the current generated by the harmonic drive on the gate.
For this, we have applied a few-MHz sine tone on the gate with varying amplitude and measured the time-averaged transmission (using technique A) of the wire as the function of the drive amplitude, $A_{\text{g}}$ and the DC bias current, $I_{\text{b}}$ at zero offset gate voltage (time-resolved measurements will be discussed later). An example is shown in Fig.~\ref{fig:fig4}c with $f_{\text{g}}=15$~MHz. At small drive amplitudes the features resemble well the undriven case, i.e. the curve presented in Fig.~\ref{fig:fig2}d with a high transmission at low DC current, and with decreasing transmission as the device switches to the N state in several steps with increasing bias current. At higher drive amplitudes the bright yellow, fully SC region shrinks and vanishes around $A_{\text{g}}\approx 0.5$~V amplitude. At the same time two better transmitting branches develop diagonally. At large amplitudes, above $A_{\text{g}}=2$~V, close to zero DC current the transmission is close to the value of the normal case, but surprisingly, by applying a finite DC bias current one finds higher transmissions in the branches.

The main features of this measurement can be captured using a simple circuit model (see results in Fig.~\ref{fig:fig4}d), where the resistance of the wire depends on the total current flowing through it and increases in a step-like fashion. The current is the sum of the DC bias and the AC displacement current component, therefore the resistance becomes time-dependent. The instantaneous transmission is calculated from the resistance with a phenomenological formula.  Finally, time-averaging for one period gives the average transmission shown in Fig.~\ref{fig:fig4}d reproducing well the experimental data. Details of the model described in the Methods. For the SC triangle at small drive amplitudes and DC bias, the total current is always below the lowest switching current, $I_{\text{c1}}$. At the tip of the triangle, the amplitude of the displacement current reaches $I_{\text{c1}}$. This feature allows for an alternative way to determine $C_{\text{eff}}$, which is presented in Supp.\ Mat.\ VI., yielding a good agreement with the one extracted from pulsed measurements on Fig.~\ref{fig:fig3}d. The diagonal, high-bias features for large amplitudes originate from the partial compensation of the displacement current with the bias current.
 
To study the time scales of the leakage-based switching mechanism close to the threshold gate voltage we return to the pulsed measurements (technique B). We suppress the DCIS by choosing a slow enough ramping of the gate voltage, such that the displacement current does not switch the wire. Then close to the threshold gate voltage we evaluate the delay of the response of the wire compared to the arrival of the gate pulse.

\begin{figure}[tb]
	\begin{center}
	\includegraphics[width=\columnwidth]{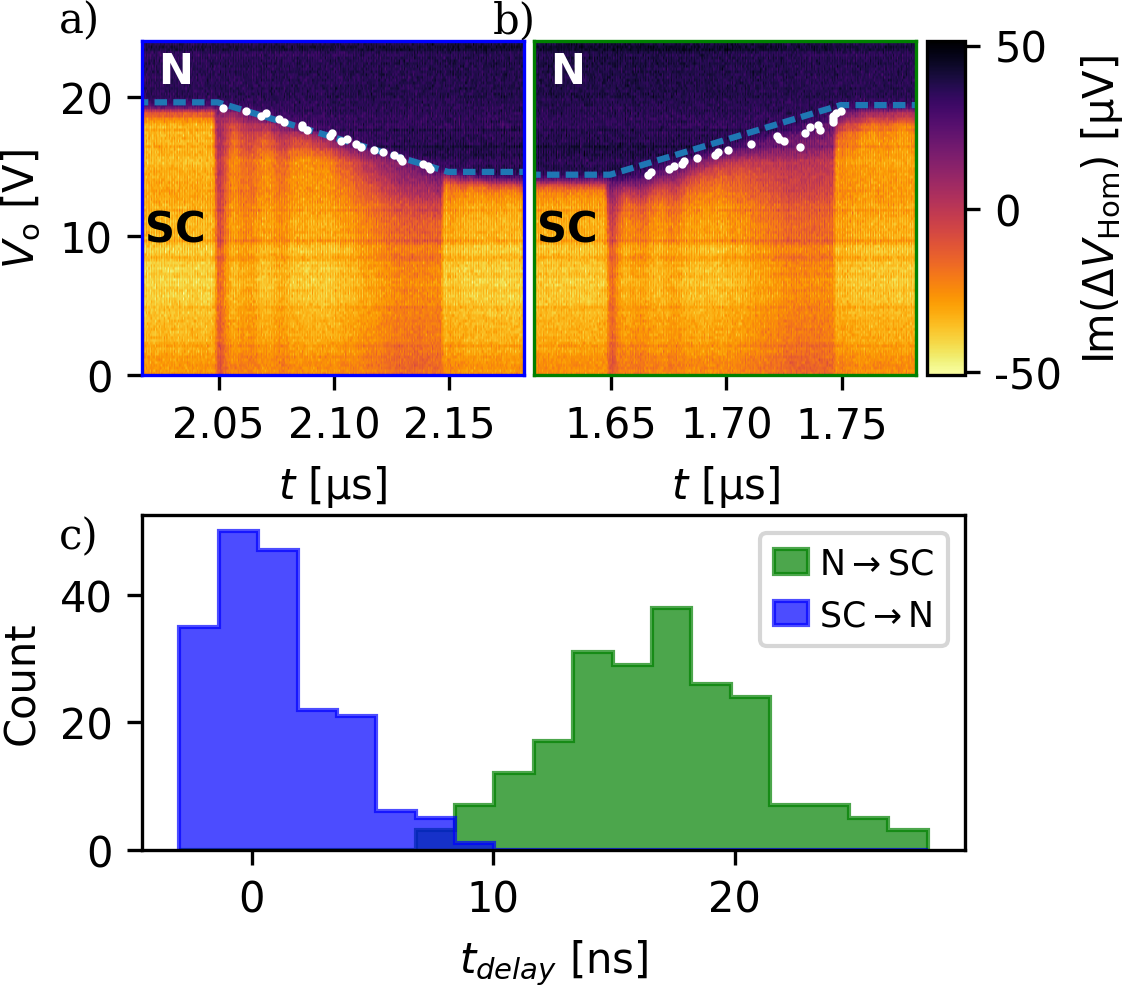}
	\caption{\textbf{Timescales of the leakage-based switching.} a) and b) zoom-in to the rising and the falling edge of the time dependence of the RF readout signal using ramped, trapezoid-like gate pulses, with 100~ns ramp time and 2.5~V pulse amplitude. Blue lines mark the $V_{\text{o}}$ value where the offset gate voltage plus the pulse is equivalent to the threshold voltage. White dots indicate the point in each time trace where the signal reaches the level corresponding to the normal state (purple) or start to decay from that level. The horizontal separation of the blue lines and the white dots correspond to the delay times of the leakage-based switchings, and plotted on panel c) as a histogram for 9 up and down steps. The switching to normal state happens in a few ns (blue), but the normal to supra transition is delayed about 15--20~ns (green).}
	\label{fig:fig5}
	\end{center}
	\end{figure}

Such a measurement is presented in Fig.~\ref{fig:fig5} using a similar, ramped pulse sequence as in Fig.~\ref{fig:fig3}a with 2.5~V amplitude and 100~ns ramp time. We focus on the ramp up and ramp down part of the pulse sequence. As the total gate voltage is given by the sum of the DC offset and the pulse, in the ramp up phase of the pulse, the DC gate threshold is decreasing. The sudden change in the signal at $t=2.05$~\textmu s and 1.65~\textmu s marks the point in time when the gate pulse reaches the wire and the shifting of the threshold starts. Using this, we can calculate, based on the total gate voltage, the actual DC threshold voltage where the SC to N transition should take place at a given time. We indicate this threshold by the blue dashed line. While the white points show the position of the transition evaluated from the high frequency transmission measurements (yellow to purple color change, for the detailed description of the evaluation see Supp.\ Mat.\ IX.). For the ramp up phase (left panel), the transition to the N state roughly coincides with the blue line. On the contrary, for the down ramp (middle panel) there is a clear delay in the signal with respect to the blue dashed line. This indicates that the switching to N state by the leakage current in almost instantaneous, but the switching back to SC is delayed. To quantify the time scales, for each time trace we identify the time delay between the expected and measured transition times and created a histogram out of the time delays (see panel c). For the ramp up, the mean value is less than 1~ns, while for the down ramp it is 19~ns. The spread of the distribution can also originate from the noise of the readout signal and the jitter of the pulse. Our measurements indicate that leakage current itself responds quite quickly to the change in the gate voltage, the wire switches to N within a few nanoseconds, but the switching back to SC state is limited, presumably due to thermal effects. The ns-switching time is one order of magnitude faster than the previously reported switching speed \cite{JointArxiv2024}. As soon as the wire switches to the N state, dissipation occurs, and due to the low phonon cooling efficiency at low temperatures, the switching back is slow.

\begin{figure}[tb]
	\begin{center}
	\includegraphics[width=\columnwidth]{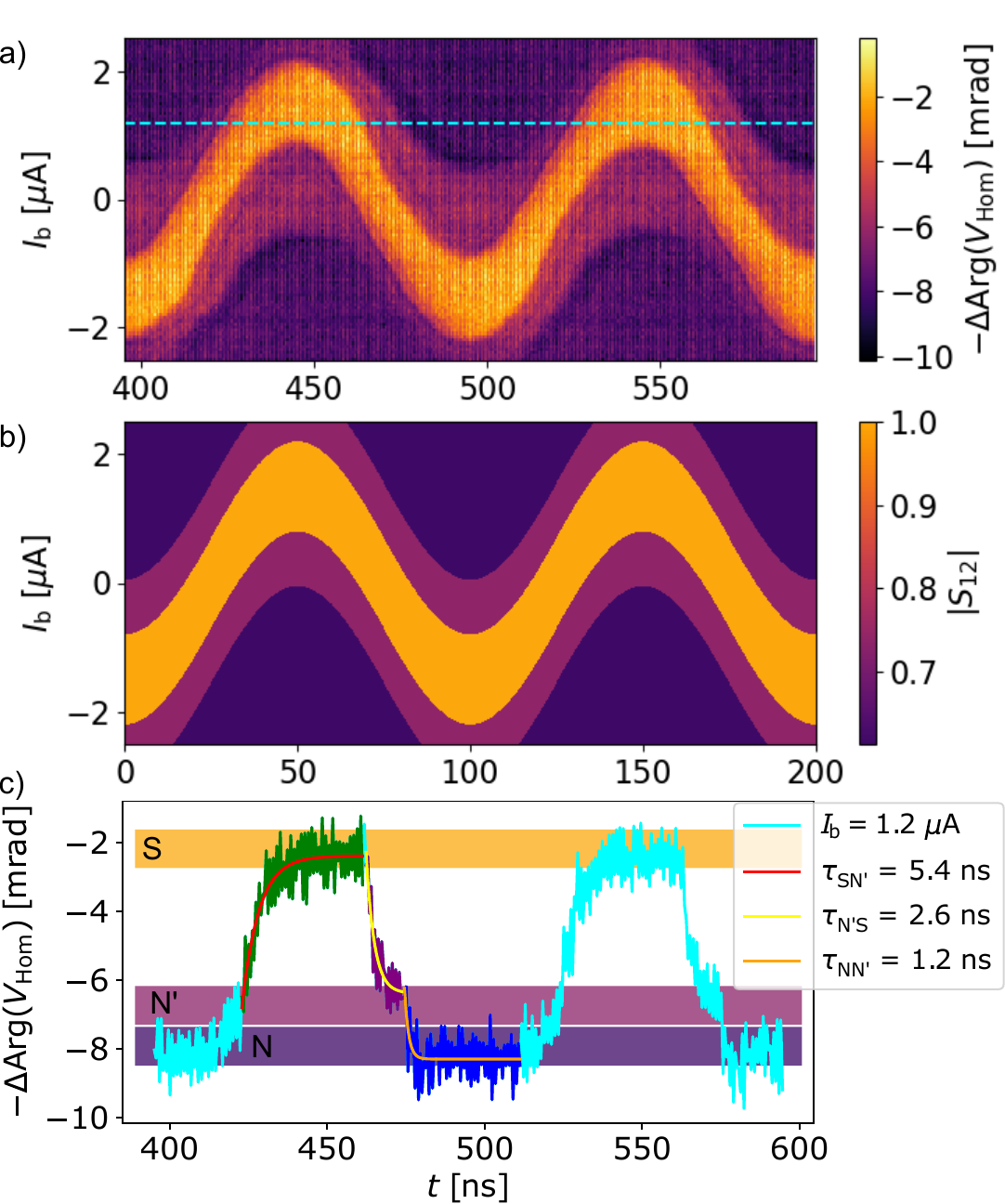}
	\caption{\textbf{Timescales of the displacement-current-based switching.} a) Time-resolved transmission (phase of the homodyne voltage) as the function of the DC bias current at zero DC gate voltage under harmonic drive using $f_{\text{g}}=10$~MHz and $A_{\text{g}}=5$~V. The boundaries of the different regions shift along the $I_{\text{b}}$ axis with the driving signal. b) Simulation of the measurement of panel a). c) Exemplary horizontal cut from panel a) at the cyan line, the color-coded sections are fitted by exponential functions yielding 5--6~ns switching time from N' to S and 1--2.5~ns for S to N' and N' to N.}
	\label{fig:fig6}
	\end{center}
	\end{figure}

Finally, we quantify the time scales corresponding to the DCIS. For this we used a harmonic driving on the gate and measured the time-resolved transmission (using technique B) as the function of the DC bias current, $I_{\text{b}}$ at zero DC offset gate voltage. An example is shown in Fig.~\ref{fig:fig6}a with $f_{\text{g}}=10$~MHz and $A_{\text{g}}=5$~V. The measurement indicates that the borders of the different regions (yellow for the SC and purple for the N state) oscillate sinusoidally with the applied gate drive. This is because at each time the sinusoidal displacement current can be compensated by the DC current, providing a zero total current and hence the wire is in the SC state. When it happens at zero displacement current (extrema of the driving signal), the SC region is centered around zero DC current, whereas when the displacement current is finite the SC region is shifted vertically. These features are nicely captured by our model (see Fig.~\ref{fig:fig6}b). In the model (detailed in the Methods) the state of the wire is only determined by the total current, which is the sum of the DC bias and the induced sinusoidal displacement current.

To evaluate the timescales we focus on an exemplary time trace of panel a) taken at $I_{\text{b}}=1.2$~\textmu A, which is plotted on panel c). The signal changes periodically between three signal levels, yellow as S, purple as N' and dark purple as N. As noted before, our device turns to normal in two steps \cite{ElalailyNanoLett2021}. The transitions between these states, visible in panel c), are fitted with exponential functions with the time constants associated to the timescales of the DCIS. Due to low resolution we did not fit the N$\rightarrow$N' transition. The N'$\rightarrow$S transition (green section) yields a timescale of $\tau_{\text{SN'}} \approx 5.4$~ns. When fitting several of these processes we obtain $3.9\pm1.3$~ns independent of the bias current (see additional examples in Supp.\ Mat.\ XI.). This is significantly faster than the leakage-based switching, which we believe can be explained by the absence of non-equilibrium phonons. With DCIS only quasiparticles are excited in the nanowire, while the leakage current generates non-equilibrium phonons as well, which also have to decay. The other switchings, the S$\rightarrow$N' ($\tau_{\text{N'S}}$, purple) and N'$\rightarrow$N ($\tau_{\text{NN'}}$, blue) yield $2.8\pm0.8$ and $1.4\pm0.5$~ns in average (on the particular plotted curves 2.6 and 1.2~ns), respectively, which is in the range of the instrumental limitations of our setup. The fast switching is also demonstrated by pulsed measurements, which are discussed in Supp.\ Mat.\ IV.
	
\section{Conclusion}

In this paper we have studied the timescales associated with the gate-based operation of an Al nanowire superconducting switch by monitoring the transmission of a $\sim 4$~GHz continuous tone through the device. The significant capacitance between the gate electrode and the current leads allows for a new type of operation, where the gate pulse switches the wire to normal state via inducing a displacement current in the wire. The switching to normal state is close to our instrumental limits, happens in about 1--3~ns, which is more than one order of magnitude faster than previously reported values \cite{RitterNatComm2021,JointArxiv2024}. The switching from the normal state back to the superconducting one is somewhat slower, 4--6~ns, presumably limited by the relaxation of the quasiparticles.

By suppressing the DCIS mechanism we also explored the timescales associated to the leakage-based switching, giving a similarly fast switching to normal state, but a slower, 15--20~ns back to the superconducting state. The latter is presumably limited by the combined thermal relaxation of the leakage-current-generated phonons and quasiparticles in the superconductor.

These timescales suggest that the displacement-current-based operation is more promising for an application than the leakage-based operation. In the DCIS operation using square pulses the device switches only during the rise or fall-time of the pulse, which could be used for fast pulse generation e.g. in neuromorphic architectures \cite{DuNatComm2017,BerggrenNanotech2020,ChenNatComm2023,NishiokaCommEng2024}. Moreover, a continuous (harmonic) driving can keep the device in normal state for arbitrarily long. For scaling up these devices, the operation should be local: the application of the gate voltage should only address a single nanowire. Whereas for the DCIS mechanism this can be well engineered by well-defined capacitors, for the leakage-based mechanism this could be a challenge for some substrates \cite{RitterNatEl2022}. The footprint of a single device could be significantly decreased by using well-designed capacitors, with high dielectric constant (e.g. a plate capacitor with $A=1$~\textmu m$^2$ area assuming $\varepsilon_{\text{r}}=25$ dielectric constant of HfO$_2$ with 10~nm thickness yields 22~fF, similar to the one in our device), whereas the crosstalk can be mitigated by using ground planes to screen the unwanted stray fields. At the moment the operation speed is limited by the relaxation, which may be improved by better thermalizing the device, or by using e.g. quasiparticle traps. Finally, for finite gate voltages, the two operation methods can be combined, where the fast switching is obtained by the DCIS mechanism, whereas the long-term memory of the state could be provided by the leakage current.

\vskip1cm
{\bf \noindent \large Acknowledgments}

This work was funded by the EU’s Horizon 2020 research and innovation program under grant SuperGate network (964398), the EIC Pathfinder Challenge grant QuKiT (101115315), by the European Research Council ERC project Twistrain, Novo Nordisk Foundation SolidQ, the COST Action CA21144 (SUPERQUMAT), OTKA K138433.
This research was supported by the Ministry of Culture and Innovation and the National Research, Development and Innovation Office within the Quantum Information National Laboratory of Hungary (Grant No. 2022-2.1.1-NL-2022-00004).
This paper was supported by the J\'anos Bolyai Research Scholarship of the Hungarian Academy of Sciences.
Supported by the EK\"OP-24-3-BME-162 and EK\"OP-24-4-II-BME-95 University Research Scholarship Program of the Ministry for Culture and Innovation from the source of the National Research, Development and Innovation Fund.
This research was supported by the Carlsberg Foundation, and the Danish National Research Foundation (DNRF 101).

\vskip1cm
{\bf \noindent \large Additional information:}
The authors declare no competing interests. Correspondence should be addressed to Z.~S..

\vskip1cm
{\bf \noindent \large Author contribution:}

T.~E. fabricated the device. 
Z.~S. and L.~K. performed the measurements.
T.~K. and J.~N. developed the nanowires.
The data analysis was performed by Z.~S., M.~K. and L.~K..
Modeling and simulation was done by Z.~S., L.~K. and G.~F. with inputs from K.~B..
The project was guided by Z.~S., P.~M. and S.~C..
The manuscript has been prepared by Z.~S., M.~K. and P.~M. with input from all authors.

\vskip1cm
{\bf \noindent \large Methods}

\vspace*{5mm}
{\bf\noindent Device and setup} 
\vspace*{2mm}

The investigated device was fabricated by depositing InAs nanowire with a 20~nm-thick full Al shell layer on an undoped Si wafer with a 290~nm-thick oxide layer. Following the nanowire deposition, four Ti/Al contacts with thicknesses of 10/80~nm and two opposite side gates made of Ti/Au with thicknesses of 7/33~nm were patterned using two separate electron beam lithography steps. In the device shown in Fig.~\ref{fig:fig1}a, the length of the nanowire segment under investigation is approximately 4.2~\textmu m.

The device was contacted in a quasi-four-point geometry. Two leads were connected to both DC and RF lines through on-board bias tees (100~$\Omega$ and 10~pF), these were used as biased leads. One of the rest of the contacts was not working, so the DC resistance was measured in a three-point configuration and the data was later corrected for line resistance of 150~$\Omega$. The presented data corresponds to the corrected values. The device was current-biased symmetrically to keep the potential of the wire constant. The RF measurement was carried out by transmitting a 3.95~GHz signal through the device applied on the current leads. The transmitted signal was downconverted by an IQ mixer using a second local oscillator and was measured by a Zurich Instruments UHFLI. Two different measurement schemes were used throughout the paper. First, the heterodyne \textit{technique A}, the transmitted signal was downconverted to a few hundred MHz and was measured by lock-in technique, where the reference signal was also generated by downconversion. In this case the time resolution was limited by integration time of the UFHLI, which has a lower limit of 30~ns, yielding to a $\sim 100$~ns time resolution. Second, the homodyne \textit{technique B}, the signal was directly downconverted to DC and was measured by the scope module of the UHFLI, allowing for a 1.8~GS/s sampling rate and the time resolution was limited by the 600~MHz bandwidth of the UHFLI. As IQ mixers were used, both quadratures of the signal was measured with both techniques. In case of technique A, the signals at the lock-in inputs are expressed as
\bean 
V_{\text{I}}(t) &=& \left|V_{\text{Het}}(t)\right| \text{cos} \left( \omega t + \text{Arg}\left(V_{\text{Het}}(t)\right)\right) \nonumber \\
V_{\text{Q}}(t) &=& \left|V_{\text{Het}}(t)\right| \text{sin} \left( \omega t + \text{Arg}\left(V_{\text{Het}}(t)\right)\right),
\eean
hence both quadratures contain both the magnitude and the phase. While for technique B they are
\bean
V_{\text{I}}(t) &=& \left|V_{\text{Hom}}(t)\right| \text{cos} \left(\text{Arg}\left(V_{\text{Hom}}(t)\right)\right) \nonumber \\
V_{\text{Q}}(t) &=& \left|V_{\text{Hom}}(t)\right| \text{sin} \left(\text{Arg}\left(V_{\text{Hom}}(t)\right)\right),
\eean
i.e. the magnitude and the phase can be determined if both quadratures are measured.
All measurements were carried out at 5 to 15 dBm readout power at top of the cryostat, and the input line had 80 dB nominal attenuation (see Supp.\ Mat.\ I.).

\vspace*{5mm}
{\bf\noindent Training effect}

Through the measurements the threshold voltage, i.e. the offset gate voltage, where the superconductivity is quenched decreased monotonically. This is due to the so-called training effect \cite{ElalailyNanoLett2021}, that is first a large gate voltage is needed to start the leakage current, but after that the current itself builds a better and better conducting channel, so lower and lower gate voltage is enough provide the same leakage current until it saturates. The DC-gated measurements here were carried out in the following chronological order: i) Fig.~\ref{fig:fig3}, with $V_{\text{th}}\approx 23$~V, ii) Fig.~\ref{fig:fig2}, with $V_{\text{th}}\approx 19$~V, iii) Fig.~\ref{fig:fig5}, with $V_{\text{th}}\approx 17$~V and iv) Fig.~\ref{fig:fig4}a and b, with $V_{\text{th}}\approx 15$~V.

\vspace*{2mm}

\vspace*{5mm}
{\bf\noindent Modeling} 
\vspace*{2mm}

Here we outline how the effect of the displacement current induced by the harmonic drive on the gate was simulated to generate Fig.~\ref{fig:fig4}d and \ref{fig:fig6}b. First, we assume two different critical currents, $I_{\text{c1}}$ and $I_{\text{c2}}$ and different constant resistance values above each critical current, $R_1$ and $R_2$, corresponding to the N' and N states, respectively. This captures the effect of having more than one steps. Second, we assume that the resistance jumps at $I_{\text{ci}}$, but otherwise it is constant, therefore
\bnen R(I)=\begin{cases} 
0 & \text{if } I < I_{\text{c1}} \\
R_1 & \text{if } I_{\text{c1}} \leq I < I_{\text{c2}} \\
R_2 & \text{if } I_{\text{c2}} \leq I \end{cases}. \eden
The current flowing through the wire is the sum a DC component and an AC one, coming from the biasing and displacement current, respectively, hence
\bnen I(t) = I_{\text{b}} + I_{\text{AC}} \text{sin}\left( \omega t \right). \eden
The AC current can also be expressed as
\bnen \label{eq:disp} I_{\text{AC}}(t) = I_{\text{disp}} = C \frac{dV_{\text{g}}(t)}{dt} = C \omega V_{\text{g}} \text{sin} \left( \omega t \right). \eden
Due to the displacement current, the resistance also becomes time-dependent. In the model we used $C=20$~fF.

Finally, we assume that the transmission instantaneously follows the resistance,
\bnen \left|S_{12}\right|(t) = \frac{474}{474+R(t)}, \eden
where the phenomenological formula introduced in Supp.\ Mat.\ III.\ is used. These steps yield the time-dependent plot in Fig.~\ref{fig:fig6}b. To generate Fig.~\ref{fig:fig4}d the transmission is time-averaged,
\bnen \left|S_{12}\right| = \left\langle \left|S_{12}\right|(t) \right\rangle, \eden
where $\left\langle \cdot\right\rangle$ denotes the averaging for one period of the gate signal.

The parameters used to generate the two figures are
\begin{table}[h!tb]
\begin{center}
\begin{tabular}{| c || c | c | c | c | c | } 
\hline
& $C_{\text{eff}}$ & $R_1$ & $R_2$ &  $I_{\text{c1}}$ & $I_{\text{c2}}$ \\
\hline
Fig.~\ref{fig:fig4}d & 20~fF & $200~\Omega$ & $300~\Omega$ & 0.9~$\mu$A & 1.8~$\mu$A  \\
\hline 
Fig.~\ref{fig:fig6}b & 20~fF & $170~\Omega$ & $300~\Omega$ & 0.7~$\mu$A & 1.55~$\mu$A  \\
\hline 
\end{tabular}
\end{center}
\end{table}

\bibliography{refs}

\end{document}